\documentclass[a4paper,twocolumn,showpacs,superscriptaddress,prl]{revtex4}
\usepackage[english]{babel}
\usepackage{graphicx}
\usepackage{xspace}

\begin{document}
\title{Transport anisotropy in biaxially strained La$_{2/3}$Ca$_{1/3}$MnO$_3$ thin films}
\author{J. Klein}
\affiliation{II. Physikalisches Institut, Universit\"{a}t zu K\"{o}ln,
Z\"{u}lpicher Str.~77, 50937 K\"{o}ln, Germany}
\author{J. B. Philipp}
\affiliation{Walther-Meissner-Institut, Bayerische Akademie der
Wissenschaften, Walther-Meissner Str.~8, 85748 Garching, Germany}
\author{G. Carbone}
\affiliation{Max-Planck-Institut f\"{u}r Metallforschung,
Heisenbergstr.~1, 70569 Stuttgart, Germany}
\author{A. Vigliante}
\affiliation{Max-Planck-Institut f\"{u}r Metallforschung,
Heisenbergstr.~1, 70569 Stuttgart, Germany}
\author{L. Alff}
\email{Lambert.Alff@wmi.badw.de}
\affiliation{Walther-Meissner-Institut,
Bayerische Akademie der Wissenschaften, Walther-Meissner Str.~8, 85748
Garching, Germany}
\author{R. Gross}
\email{Rudolf.Gross@wmi.badw.de}
\affiliation{Walther-Meissner-Institut,
Bayerische Akademie der Wissenschaften, Walther-Meissner Str.~8, 85748
Garching, Germany}

\date{received June 04, 2002}

\begin{abstract}
Due to the complex interplay of magnetic, structural, electronic,
and orbital degrees of freedom, biaxial strain is known to play an
essential role in the doped manganites. For coherently strained
La$_{2/3}$Ca$_{1/3}$MnO$_3$ thin films grown on SrTiO$_3$
substrates, we measured the magnetotransport properties both
parallel and perpendicular to the substrate and found an anomaly
of the electrical transport properties. Whereas metallic behavior
is found within the plane of biaxial strain, for transport
perpendicular to this plane an insulating behavior and non-linear
current-voltage characteristics (IVCs) are observed. The most
natural explanation of this anisotropy is a strain induced
transition from an orbitally disordered ferromagnetic state to an
orbitally ordered state associated with antiferromagnetic stacking
of ferromagnetic manganese oxide planes.
\end{abstract}

\maketitle

\vspace*{-10cm}\noindent
 to appear in {\em Physical Review B} (Brief
 report)

\vspace*{9.5cm}

It is well known that the physics of the doped perovskite
manganites is determined by a complex interplay of structural,
magnetic, electronic, and orbital degrees of freedom. While the
classical double exchange model can qualitatively explain the
transition from a paramagnetic insulating to a ferromagnetic
metallic state \cite{Zener:51}, for a more complete understanding
of the physics of the manganites electron-lattice coupling has to
be included \cite{Millis:95}. Recently, Millis {\em et al.} have
pointed out that uniform compression, as realized by hydrostatic
pressure, increases the electron hopping amplitude favoring a
ferromagnetic metallic state \cite{Millis:98}. In contrast,
biaxial strain, as realized in epitaxial thin films grown on
substrates with significant lattice mismatch, enhances the
Jahn-Teller distortions favoring an insulating state due to the
tendency of the electrons to become localized \cite{Millis:98}.
Fang {\em et al.~}have calculated the phase diagram of the almost
tetragonal doped manganite La$_{1-x}$Sr$_{x}$MnO$_3$ as a function
of biaxial strain by studying the instabilities of the
ferromagnetic state \cite{Fang:00}. Their results are in ageement
with experiments on biaxially strained La$_{1-x}$Sr$_{x}$MnO$_3$
thin films \cite{Konishi:99}. In their work, it has been shown
that the orbitally disordered ferromagnetic state (F) is unstable
against orbital ordered states with layer (A) and chain (C) type
antiferromagnetism. In turn, it is expected that these different
magnetic states are associated with different magnetotransport
behavior via the double exchange mechanism. The further
investigation of the validity of the strain phase diagram, the
corresponding magnetotransport properties, and the extendability
to doped manganites with strong tilt of MnO$_6$ octahedra
\cite{Vigliante:2001} and phenomena as charge ordering as for
example La$_{1-x}$Ca$_{x}$MnO$_3$ is of great interest to gain
more insight into the physics of these materials and its
dependence on lattice distortions. For the purpose of this study,
it is important to verify the coherency of the strained state of
the doped manganite, in order to be able to determine properly the
intrinsic properties within the biaxial strain phase diagram, and
to exclude effects of nonuniform strain distribution or relaxation
effects \cite{Biswas:00,Biswas:01}.

In this Letter we present a careful study of the structural,
electronic, and magnetic properties of coherently strained
La$_{2/3}$Ca$_{1/3}$MnO$_3$ (LCMO) thin films and
LCMO-La$_{2/3}$Ba$_{1/3}$MnO$_3$ (LBMO) heterostructures on
SrTiO$_3$ substrates. In addition to previous studies, transport
properties have been measured both parallel and perpendicular to
the plane of biaxial strain. The key result of our study is that
biaxial strain results in highly anisotropic transport properties:
Whereas insulating behavior and non-linear IVCs are observed
perpendicular to the biaxially strained plane (parallel to the $c$
axis), the in-plane transport is metallic below the Curie
temperature $T_C$. The saturation magnetization of biaxially
strained LCMO is strongly reduced compared to the bulk material or
the less strained LBMO films. It is shown that this behavior is
not due to interface effects between different layers
\cite{Bibes:01}, but is an {\em intrinsic} property of the
biaxially strained LCMO arising most likely from a strain induced
orbital ordering. Another important result is that in strained
LCMO a low-resistance state can be induced by applying either a
magnetic field or a high current density. A similar behavior is
for example observed in Pr$_{0.7}$Ca$_{0.3}$MnO$_3$ and
Nd$_{0.5}$Sr$_{0.5}$MnO$_3$ and has been attributed to a magnetic
field or current induced (local) melting of a charge
resp.~orbitally ordered ground state
\cite{Asamitsu:97,Fiebig:98,Guha:00a}.

We have grown LCMO films and LBMO-LCMO-LBMO heterostructures on
SrTiO$_3$ ($a \simeq 3.905$\,{\AA}) substrates using pulsed laser
deposition \cite{Gross:2000a,Klein:2001a}. The lattice mismatch
between the SrTiO$_3$ substrate and LCMO ($a_{\rm bulk} \simeq
3.864$\,{\AA} in pseudocubic notation) is -1.2\% resulting in
in-plane tensile strain, whereas the mismatch between SrTiO$_3$
and LBMO ($a_{\rm bulk}\simeq 3.910$\,{\AA}) is only about 0.13\%
resulting in very small compressive strain. Note that in the
LBMO-LCMO-LBMO heterostructures the LBMO layers provide low
resistance contacts to the (ultra)thin LCMO films allowing for a
homogeneous current feed for transport perpendicular to the film.
Furthermore, effects of a possible surface \lq\lq dead layer"
\cite{Sun:99} are avoided in the multilayer structure.
Layer-by-layer growth of the films was confirmed by a high
pressure reflection high energy electron diffraction (RHEED)
system showing clear growth oscillations
\cite{Rijnders:97,Klein:00}.

As has already been stressed it is important to prove the
coherency of the strained state. The coherent film thickness was
determined from Laue oscillations in $\theta-2\theta$ x-ray scans.
It was found that both LCMO and LBMO films grow coherently
strained on SrTiO$_3$ substrates up to a thickness of at least
60\,nm, consistent with literature \cite{Bibes:01,Izumi:98}. The
out-of-plane ($c$ axis) and in-plane ($a$ axis) lattice parameters
of the LCMO thin films grown on SrTiO$_3$ have been determined
from the (002) and (103) reflections. The in-plane film lattice
parameters coincide throughout the whole layer with the substrate
lattice parameters. That is, the in-plane lattice parameter of
LCMO is enlarged, while the out-of-plane lattice constant is
reduced, leading to a tetragonal lattice distortion with
$c/a\approx0.985$. The tetragonal distortion can be viewed as a
Jahn-Teller like distortion resulting in an increased Jahn-Teller
splitting of the Mn $e_g$ levels and, in turn, in a tendency of
the electrons to become localized.

\begin{figure}[tb]
\centering{%
\includegraphics
[width=0.95\columnwidth,clip=]{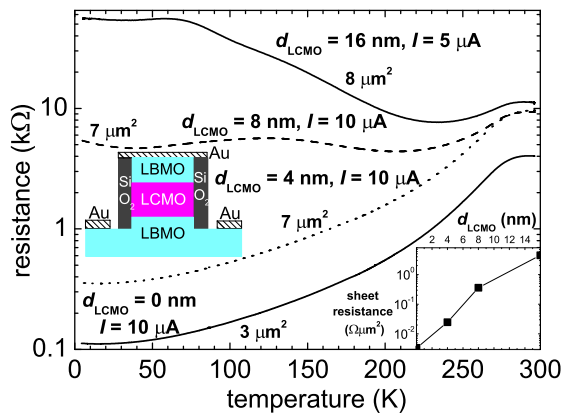}}
\vspace*{0mm}\\
\caption{Resistance vs temperature curves of mesa type
LBMO-LCMO-LBMO heterostructures as sketched in the inset.  The
thickness $d_{\rm LCMO}$ of the LCMO layer, the size of the mesa,
and the measurement current is given next to the corresponding
curves. The inset shows the sheet resistance vs $d_{\rm LCMO}$ at
10\,K.}
 \label{rt}
\end{figure}

\begin{figure}[tbh]
\centering{%
\includegraphics
[width=0.95\columnwidth,clip=]{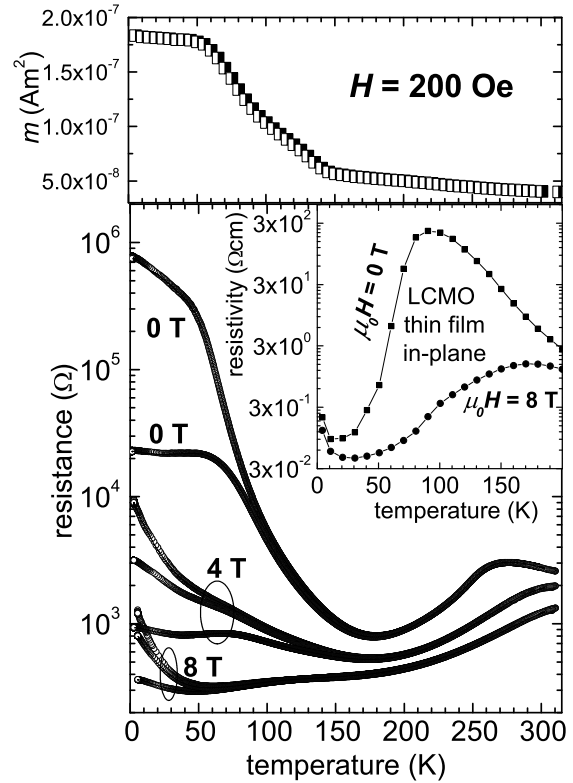}}
\vspace*{0mm}\\
\caption{Resistance vs temperature curves of a LBMO-LCMO-LBMO
($d_{\rm LCMO}=7.2$\,nm) heterostructure measured for current
perpendicular to plane at different applied magnetic fields and
currents (0\,T: 0.1 and 10\,$\mu$A; 4\,T: 0.1, 10 and 100\,$\mu$A;
8\,T: 0.1, 10 and 100\,$\mu$A, the resistance decreases with
increasing current). The sample configuration is shown in the
inset of Fig.~\ref{iuhy}. The inset shows the $\rho(T)$ curves of
a LCMO film for current applied in-plane. In the upper picture the
magnetization $m$ is shown.}
 \label{rthi}
\end{figure}

In order to measure the electrical transport properties
perpendicular to the film plane, mesa structures (see inset of
Fig.~\ref{rt}) were patterned using optical lithography and Ar ion
beam etching. The mesas have typical area of several $\mu$m$^2$.
Fig.~\ref{rt} shows the resistance vs temperature, $R(T)$, curves
for LBMO-LCMO-LBMO mesa structures with different thickness
$d_{\rm LCMO}$ of the LCMO layer. For comparison, a sample without
LCMO layer ($d_{\rm LCMO}=0$) is shown. It is evident that the
resistance increases with increasing $d_{\rm LCMO}$. We estimate
the involved resistivities $\rho_c$ to about several $\Omega$m at
10\,K. Note that due to the non-linear IVCs, it is difficult to
obtain a meaningful $\rho_c$ or thickness dependence $\rho_c(d)$.
In the inset we show the sheet resistance as a function of $d_{\rm
LCMO}$. The strong increase of resistance with increasing layer
thickness can partially be due to tunneling through the barrier.
The important point is that the resistivity does not show any
tendency to saturation meaning the resistance is intrinsic to the
barrier and not to the interface. Furthermore, the metallic $R(T)$
behavior below $T_C$ that is observed for the sample with $d_{\rm
LCMO}=0$ turns into an semiconducting or insulating $R(T)$
behavior with increasing $d_{\rm LCMO}$. The low-temperature
plateau (below 60\,K) of the resistivity corresponds to a plateau
in the temperature dependence of the magnetization (see upper
panel of Fig.~\ref{rthi}) as expected for double exchange
materials. This thickness dependence clearly shows that the
insulating $R(T)$ behavior is {\em not} due to the patterning
process, due to a contact resistance between the gold contact
layer and the manganite film, nor due to interface effects. One
can conclude that the insulating behavior for transport along the
$c$ axis observed at low temperatures is an {\em intrinsic}
property of the coherently strained LCMO thin films. We note that
in the work of Bibes {\em et al.} interfaces between magnetic LCMO
and non-magnetic SrTiO$_3$ are shown to favor phase segregation
\cite{Bibes:01}. Inhomogeneous transport properties were found in
La$_{0.7}$Ca$_{0.3}$MnO$_3$/SrTiO$_3$ heterostructures and
interpreted as arising from magnetic interface disorder
\cite{Jo:99}. In our case, the interfaces are between two
differently doped manganites. While it remains to be investigated
whether interdiffusion is present at such interfaces, the effects
observed here are related to the magnetic order of the whole thin
film layer.

\begin{figure}[tb]
\centering{%
\includegraphics
[width=0.95\columnwidth,clip=]{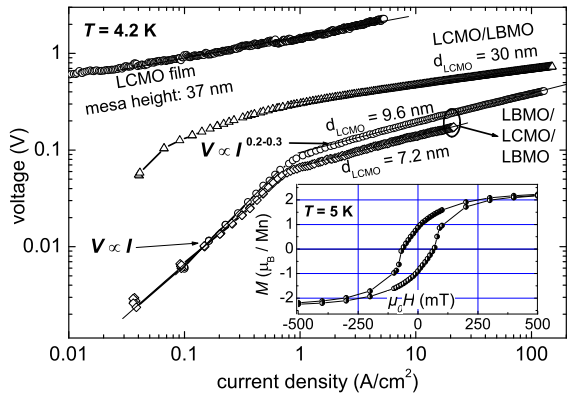}}
\vspace*{0mm}\\
\caption{IVCs measured along the $c$ axis direction at 4.2\,K for
two LBMO-LCMO-LBMO mesa structures with $d_{\rm LCMO}=7.2$\,nm
and 9.6\,nm. Also shown are the IVCs for mesa structures patterned
into a LBMO-LCMO bilayer with $d_{\rm LCMO}=30$\,nm and a single
LCMO with a mesa height of 37\,nm. The inset shows the
magnetization vs applied field curve of a $57.5$\,nm thick LCMO
film at 5\,K.}
 \label{iu}
\end{figure}

In Fig.~\ref{iu} we show the IVCs of several mesa structures with
different values of $d_{\rm LCMO}$ measured in $c$ axis direction,
i.~e.~with current perpendicular to the biaxially strained plane.
While for low enough current densities $V\propto I$, all mesas
show highly non-linear IVCs for higher current densities following
a $V\propto I^{n}$ dependence with $n\approx0.2-0.3$. In contrast,
the IVCs measured with the current in-plane are ohmic. For
comparison, mesa structures have been patterned into single LBMO
films ($d_{\rm LCMO}=0$) that are well lattice matched to the
SrTiO$_3$ substrate. For these samples, ohmic IVCs have been
obtained both for the current applied in- {\em and} out-of-plane.
From this and the fact that the measured voltage clearly increases
with increasing $d_{\rm LCMO}$ (see discussion before),
preparation or interface effects can be excluded as the origin of
the nonlinear IVCs. Hence, we can conclude that for current
perpendicular to the plane of biaxial strain the insulating
behavior of the coherently strained LCMO films is associated with
non-linear current transport. We note that the non-linearity
becomes smaller with increasing $T$ and vanishes at the $T_C$ of
the LCMO thin films ranging between 100 and 150\,K. $T_C$ was
determined from magnetization measurements of the bi/trilayer
films before mesa patterning. This strongly suggests that the
electronic anisotropy is coupled to the magnetic properties of the
LCMO films.

In agreement with previous experiments we found a strain
dependence of the saturation magnetization $M_S$ of the LCMO
films. Zandbergen {\em et al.} reported $M_S\simeq 2.5\,\mu_B$/Mn
atom at 5\,K for a 6\,nm thick, coherently strained LCMO film on
SrTiO$_3$ \cite{Zandbergen:99}. Consistently, we have measured a
value of $M_S\simeq 2.2\,\mu_B$/Mn atom at 5\,K for a 57.5\,nm
thick LCMO film (see inset of Fig.~\ref{iu}). In contrast, almost
strain free LBMO films on SrTiO$_3$ show the expected saturation
magnetization of about $3.67\,\mu_B$/Mn atom. These findings are
in agreement with the phase diagram predicted by Fang {\em et
al.}~\cite{Fang:00} where for tensile strain an instability to an
A-type antiferromagnetic state is predicted. Another interesting
observation is that the size of the hysteresis loop in $M(H)$
curves also depends on strain. While for almost strain free LBMO
films on SrTiO$_3$ a coercive field of $\mu_0H_c \simeq 10$\,mT is
observed at 5\,K, a much larger value of $\mu_0H_c\simeq 70$\,mT
is measured for the strained LCMO films.

Fig.~\ref{rthi} gives an overview on the field, current and
temperature dependence of the resistance and magnetization in
strained LCMO films by showing $R(T)$ curves recorded at different
applied fields and currents. Applying high magnetic fields results
in a strong suppression of the resistance at all $T$.  Also, due
to the non-linearity of the IVCs the measured resistance is
reduced when the applied current is increased below about 150\,K.
We note that for $T\lesssim 150$\,K the resistance is dominated by
the LCMO layer, only around 250\,K it is dominated by the LBMO
layer. Fig.~\ref{rthi} clearly shows that with increasing field,
the non-linearity becomes weaker and also the onset temperature of
the non-linear behavior is shifted from about 100\,K at 0\,T to
50\,K at 8\,T. For comparison, the inset of Fig.~\ref{rthi} shows
the in-plane $\rho(T)$ curves of a LCMO film ($d_{\rm
LCMO}=57.5$\,nm). Clearly, a metallic $\rho(T)$ behavior is
observed below the peak temperature. The same film shows an
insulating $R(T)$ behavior for current perpendicular to plane.

\begin{figure}[tb]
\centering{%
\includegraphics
[width=0.95\columnwidth,clip=]{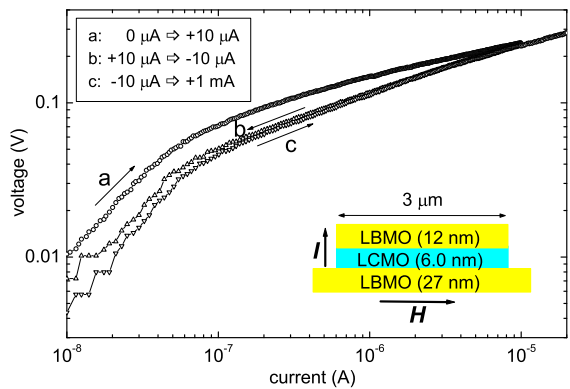}}
\vspace*{0mm}\\
\caption{IVCs of a LBMO-LCMO-LBMO mesa structure ($d_{\rm
LCMO}=6.0$\,nm) measured for current perpendicular to plane at
5\,K and zero field. The different curves have been obtained in
successive current sweeps with increasing amplitude. The mesa
area was $9\,\mu$m$^2$.}
 \label{iuhy}
\end{figure}

To further study the effect of the applied current, we have
measured IVCs for successive current sweeps with increasing
amplitude as shown in Fig.~\ref{iuhy}. After zero field cooling,
the current is first increased to 10\,$\mu$A (curve a)
corresponding to a current density of about 100\,A/cm$^2$. On
decreasing the current again (curve b), a lower voltage is
measured at the same current values, i.e. the applied current of
10\,$\mu$A has switched the sample to a state with lower
resistance. Applying a high magnetic field (8\,T) has the same
effect as applying a high current (1\,mA). After applying a field
of 8\,T, the measured IVC are stable and the resistance is
independent on the applied current. Thus, both a high magnetic
field and a high current density induce a state with reduced
resistivity.

We now address the possibility of a phase separated state in fully
strained LCMO. We note that in the case of inhomogeneously
strained or relaxed films (e.g. due to island growth), it is
plausible to assume a phase separated state with ferromagnetically
and antiferromagnetically ordered clusters, as has been discussed
recently for LCMO thin films grown on LaAlO$_3$
\cite{Biswas:00,Biswas:01}. A similar phenomenon has been observed
for bulk Pr$_{0.7}$Ca$_{0.3}$MnO$_3$ samples \cite{Fiebig:98} as
well as for LCMO bulk and thin film samples \cite{Faeth:99}.
However, phase separation cannot explain the transport {\em
anisotropy} present in our samples, but would be expected to lead
to direction {\em in}dependent behavior.

All our experimental observation can be naturally described by the
assumption of strain induced orbital order as predicted by Fang
{\em et al.}~\cite{Fang:00}. For tensile strain ($c/a < 1$, as
present in our samples), a transition from the conventional double
exchange mediated, orbital disordered F state to the orbital
ordered A state, which is composed mainly by $d_{x^2-y^2}$ states,
is expected. Whereas in the F state the spins are aligned parallel
in adjacent planes, in the A state anti-parallel orientation of
the ferromagnetically ordered planes is present. That is, the
gradual transition from the F to the A state is accompanied by a
strong reduction of saturation magnetization in agreement with the
strongly reduced value obtained in our experiments. Furthermore,
the A-type antiferromagnetic state can be metallic only within the
ferromagnetically ordered plane, but is insulating in
perpendicular direction. This again is in agreement with our
experimental observation. That is our data can be interpreted by a
strain induced orbital ordering effect at fixed doping. Evidently,
a sufficiently high current density allows to switch between the
competing states with high current density or high field favoring
the F state which can be interpreted as (local) melting of the
orbital order by introduction of highly spin-polarized carriers.

In summary, we have investigated coherently strained LCMO films.
The biaxial strain was found to induce anisotropic transport
properties at low temperatures with metallic and insulating
behavior for current in- and out-of-plane. It has been shown that
this behavior is not due to inhomogeneous interface effects or
phase separation. We suggest strain induced orbital ordering as
the origin of the observed behavior in agreement with theoretical
predictions. We also have shown that by applying a high current
density, a low-resistance state can be induced. This effect may be
of interest with respect to magnetoelectronic devices.

This work was supported by the Deutsche Forschungsgemeinschaft.
One of us, J. K., acknowledges support by the Graduiertenkolleg
549.

\end{document}